\documentclass[pdflatex,sn-aps]{sn-jnl}

\usepackage{graphicx}%
\usepackage{multirow}%
\usepackage{amsmath,amssymb,amsfonts}%
\usepackage{amsthm}%
\usepackage{mathrsfs}%
\usepackage[title]{appendix}%
\usepackage{xcolor}%
\usepackage{textcomp}%
\usepackage{manyfoot}%
\usepackage{booktabs}%
\usepackage{listings}%
\usepackage{hyperref}
\usepackage{geometry}
\theoremstyle{thmstyleone}%
\theoremstyle{thmstyletwo}%

\theoremstyle{thmstylethree}%

\begin{document}

\title[Article Title]{\textbf{Phase space analysis of an exponential model in $f(Q)$ gravity including linear dark-sector interactions}}


\author*[1]{\fnm{Ivan R.} \sur{Vasquez}}\email{itsivanvasquez@gmail.com}

\author[1]{\fnm{A.} \sur{Oliveros}}\email{alexanderoliveros@mail.uniatlantico.edu.co}

\affil*[1]{\orgdiv{Grupo de Física de Partículas Elementales y Cosmología, Programa de Física}, \orgname{Universidad del Atlántico}, \orgaddress{\street{Carrera 30 No. 8-49}, \city{ Puerto Colombia}, \state{Atlántico}, \country{Colombia}}}


\abstract{We present a cosmological analysis of an exponential $f(Q)$ gravity model, within the dynamical systems formalism. Following the method introduced by Böhmer \textit{et al} [Universe \textbf{9} no.4, 166 (2023)], the modified Friedmann modified equations are successfully reduced to an autonomous system. Given the exponential form of $f(Q)$, the equilibrium conditions result in transcendental equations, which we approximate to identify the critical points. We therefore perform a general stability analysis of these points in terms of the model parameters. Finally, we extend the model by including a linear dark energy-dark matter interaction, where the equilibrium points are found with their stability properties. The model exhibits the three main domination epochs in the Universe, as well as a non-trivial impact on the late-time de Sitter attractor.}





\maketitle

\section{Introduction}\label{sec1}

The current paradigm in cosmology states that the universe is dominated at present epochs by a fluid with an equation of state of the form $\rho/p=-1$, coming from the cosmological constant term $\Lambda$, introduced by Einstein \cite{Einstein1, Einstein2} in the 20th century. $\Lambda$, along with baryonic matter, dark matter, and radiation, constitutes the so-called $\Lambda$CDM model \cite{lambdaterm}. This model has been heavily preferred by multiple observations \cite{Planck2018, Riess, supernova1, supernova2,Brout}. However, recently the Dark Energy Spectrographic Imaging (DESI) collaboration revealed a statistically significant probability that the universe is dominated by a dynamical fluid, contrary to the $\Lambda$CDM paradigm \cite{desi, desi2, desi3}. These observations open up a fresh perspective for dynamical models of dark energy, which substitute the cosmological constant term. Over the years, a great number of proposals have been introduced, mainly exploiting the Einstein-Hilbert action, either by modifying the energy-momentum tensor $T_{\mu\nu}$ or by modifying the geometry of the Einstein tensor $G_{\mu\nu}$ (for a review, see \cite{copeland, Bamba, peebles} and references therein). This is called modified gravity.

We are interested in metric-affine theories, constructed as equivalent alternatives to Einstein's General Relativity. In these theories, the Ricci curvature is replaced by geometrical quantities called torsion and non-metricity (see these works for a complete review on the subject \cite{fqreview, trinity, jackson}). In the same manner as the successful $f(R)$, these theories may exhibit extended versions that appear naturally such as: $f(T)$ (Extended Teleparallel Gravity), $f(Q)$ (Extended Symmetric Teleparallel Gravity) \cite{coincident}. In this sense we are particularly focused in $f(Q)$ theories, as they offer attractive perspectives on phenomenology, applications that include cosmology. Over the years, a growing number of models have been proposed in the $f(Q)$ framework, and a large volume of work has been derived from such models. These works span models equivalent to GR in the coincident gauge, which show deviant behaviors in perturbative regimes \cite{signatures, cosmologyfq}. In the case of other models, we have power law models \cite{MandalQ, narawade, ayuso} in which parameter constraints are performed, as well as studies of perturbations. Some other works may include the following \cite{sahlu, Myrzakulov, Odintsov1, Odintsov2, koussour}, as well as recent additions in terms of logarithmic models \cite{sudharani, karmakar}. In the case of exponential models, Anagnostopoulos \textit{ et al.}, probed a $\Lambda$-free model in the late-time universe and in BBN constraints \cite{anagnos, anagnos2}. These exponential models have been explored in other works of modified gravity, such as \cite{Cognola, Odintsov, Granda, OliverosAcero2}. A specific variation of the exponential model with a limit to $\Lambda$CDM was recently worked out in the case of $f(Q)$ theories \cite{OliverosAcero, vasquezoliv}, which has been parametrically constrained using $H(z)$ datasets, analyzed both at background and scalar perturbative levels, and studied its impact in large-scale structures.

Every proposed model should not only account for the observed late-time accelerated expansion, but they should also describe every epoch of fluid domination in the history of the Universe. A usual tool to study this aspect is the dynamical systems formalism, which in cosmology has proven to be very useful (see \cite{bahamondedynamical, wainwright} for a complete summary and references therein). In the case of modified gravity theories, such as $f(R), f(T), f(Q)$ and others, Böhmer \textit{et al.} have found a strategy that allows the models in these theories to be worked out as autonomous differential equations \cite{boehmermod, boehmerfq}. Many interesting results have been obtained in this regard for $f(Q)$ theories, see---\cite{khyllepdynam, khyllep2, shabani, andronikos, shabani2, dutta}.  In this work, we apply the dynamical system formalism to the previously introduced model in $f(Q)$ cosmology found in \cite{OliverosAcero, vasquezoliv}, whose main feature is to be a smooth perturbative model around $\Lambda$CDM, which provides a viable model since it has been analytically analyzed and statistically constrained with $H(z)$ datasets. 

The paper is structured as follows: Section \ref{dynamical1} is intended to introduce the foundations of $f(Q)$ gravity to the reader, along with preliminary concepts from dynamical systems. In Subsection \ref{bohmer}, the modified Friedmann equations are worked out as an autonomous dynamical system, and the general analysis of the resulting equations is shown. In section \ref{phasespace}, the critical points are obtained and characterized, as well as phase space portraits and a linear stability analysis. Section \ref{dmdeinterac} serves as a general discussion on interacting DE-DM models and study of the impact of the interacting term in the evolution of DM and DE densities. Finally, in section \ref{interactingde} the model is extended to include a linear dark-sector interaction, where we find critical points, phase space portraits, and perform linear stability analysis on the extended model. Section \ref{conclusions} contains the main discussion of the results and conclusions.

\section{Dynamical system formulation for exponential $f(Q)$ model}\label{dynamical1}
\noindent Generally, the action for an $f(Q)$ gravity model in the presence of matter is of the form:
\begin{equation}\label{eq_action}
\mathcal{S}[g]=\frac{1}{\kappa}\int_{\mathcal{M}}d^{4}x\sqrt{-g}\left(f(Q)+\kappa \mathcal{L}_{m}\right),
\end{equation}
where $g$ denotes the determinant of the metric tensor, $g^{\mu\nu}$, with $\kappa=16\pi G$ (from now on, we will use geometrized units $G=c=1$), and $\mathcal{L}_{matter}$ is the lagrangian of matter fields. The term $f(Q)$ is for now an arbitrary functional of the non-metricity scalar $Q$, and the integral is performed over all spacetime $\mathcal{M}$. Note that the action is a functional of the metric $g$ only, this is possible by working in the coincident gauge for which the connection $\Gamma$ is set to vanish \cite{fqreview, coincident}. For the coincident gauge, the non-metricity tensor reduces to a partial derivative of the metric tensor,
\begin{align}
    Q_{\beta\mu\nu}=\partial_{\beta}g_{\mu\nu}.
\end{align}
Beyond that, the usual non-metricity scalar can be written in terms of the Levi-Civita connection, thus it enters the action entirely depending on the metric. Given by
\begin{align}
    Q\rightarrow g^{\mu\nu}(\gamma^{\alpha}_{\alpha\beta}\gamma^{\beta}_{\mu\nu}-\gamma^{\alpha}_{\beta\mu}\gamma^{\beta}_{\nu\alpha}),
\end{align}
where we denote the Levi-Civita connection with $\gamma$, here the action with a Lagrangian $\mathcal{L}_{CGR}[g]=Q$ is equivalent to the Einstein-Hilbert action without a boundary term. The field equations (variations w.r.t the metric) and connection field equations (variations w.r.t the connection) result in,
\begin{align}
    \frac{2}{\sqrt{-g}}\nabla_{\alpha}\left[\sqrt{-g}f'(Q)P^{\alpha}_{\mu\nu} \right]+f'(Q) q_{\mu\nu}-\frac{1}{2}f(Q)g_{\mu\nu}:=\mathcal{M}_{\mu\nu}, \\
    \nabla_{\mu}\nabla_{\nu}(\sqrt{-g}P^{\mu\nu}_{\alpha})=\mathcal{C}_{\alpha}.
\end{align}
As casted in \cite{fqreview}, $P^{\alpha}_{\mu\nu}$ is the non-metricity conjugate defined via $Q=P_{\alpha\mu\nu}Q^{\alpha\mu\nu}$, $q_{\mu\nu}=\partial_{g^{\mu\nu}}Q$ and the operators $\nabla_{\alpha}$ is a $\Gamma$-general covariant derivative. However, the identity $\mathcal{D}_{\mu}\mathcal{M}^{\mu}_{\nu}=0$ is not satisfied within this formulation, leaving that the expressions $\mathcal{C}_{\nu}$ do not lead to mere identities, but dynamical equations adding extra degrees of freedom propagated in the theory. Even though, one may deal with extra degrees of freedom, $f(Q)$ can also be formulated with nontrivial connections, applied to cosmological contexts and find enrichful models within \cite{guzman, shi, ayuso2}.  In the flat Friedman-Robertson-Walker (FRW), the metric is given by:
\begin{equation}\label{eq_FRWmetric}
ds^2 = -dt^2 + a^2(t)\delta_{ij}dx^idx^j,
\end{equation}
with $a(t)$ representing the scale factor, and the energy-momentum tensor $T_{\mu\nu}=-2\frac{\delta(\sqrt{-g}\mathcal{L}_{m})}{\delta g^{\mu\nu}}$ chosen to be the perfect fluid form $T_{\mu\nu}=(p+\rho)g_{\mu\nu}+p_{g\mu\nu}$ in the cosmological context. Then, the resulting field equations are of the form:
\begin{equation}\label{eq_timeEq}
6f'H^2 - \frac{1}{2}f = 8\pi\rho,
\end{equation}
and 
\begin{equation}\label{eq_spaceEq}
(12H^2f'' + f')\dot{H} = -4\pi(\rho+p),
\end{equation}
where $\rho$ is the total energy density of the fluids present in the universe. Here $f' = \frac{df}{dQ}$, $f''=\frac{d^{2}f}{dQ^2}$, and the over-dot denotes a derivative with respect to cosmic time $t$. Therefore, $H\equiv \dot{a}/a$ is the Hubble parameter. Further, the non-metricity scalar associated with the metric (\ref{eq_FRWmetric}) in the coincident gauge is given by, 
\begin{equation}\label{eq_Qscalar}
Q=6H^2.
\end{equation}
It should be noted that, in this scenario, the standard Friedmann equations of General Relativity (GR) plus the cosmological constant are recovered when assuming $f(Q)=Q+2\Lambda$. 

In general, dynamical systems formalism deals with systems that evolve along a background parameter, such as time. In this way, the state of a system can be represented in a given instant of time by an element $\textbf{x}\in X$; where $X$ is a \textit{state space} or \textit{phase space}, which is usually finite-dimensional as $\mathbb{R}^{n}$. Therefore, the system is described by a set of differential equations on $X$ in autonomous form
\begin{align}
    \dot{\textbf{x}}=\textbf{f}(\textbf{x}); \quad \dot{\textbf{x}}=\frac{d\textbf{x}}{dt}.
\end{align}
Here $\bf{f}(x)$ is a set of functions such as, $f_{i}: X\rightarrow X; \forall f_{i}\in \bf{f}$. In order to bring our model into the dynamical system formalism, let us consider Friedmann equations (\ref{eq_timeEq}, \ref{eq_spaceEq}) by rewriting $f(Q)=Q+F(Q)$, where the function $F(Q)$ account for non-linear modifications to the action. In this form we get,
\begin{align}\label{fried1}
    3H^2+QF'(Q)-\frac{1}{2}F'(Q)=8\pi \rho.
\end{align}
We then normalize the equation by dividing $3H^2$, therefore:
\begin{align}\label{norm1}
    \frac{8\pi \rho}{3H^2}+\frac{\frac{1}{2}F'(Q)-QF(Q)}{3H^2}=1.
\end{align}
This form is general as obtained in \cite{khyllepdynam}, where $F'(Q)$ again represents the derivative $d_{Q}F$. The second Friedmann equation is obtained by taking the time derivative in (\ref{fried1}), and using continuity equation $\dot{\rho}=-3H(p+\rho)$:
\begin{align}\label{friedF}
(2QF''+F'+1)\dot{H}=-4\pi(p+\rho),
\end{align}
where we have used $Q=6H^2$. Proposing from this point a set of variables that will allow us to write the equations as an autonomous system is usually challenging, as there is no unique method \cite{bahamondedynamical}, especially in modified gravity models. In the standard method, one simply chooses variables that normalize equation (\ref{norm1}); this method turns out to be unfeasible because the dependence of the variables on $H^2$ persists, so we will opt to apply the method used by Böhmer \textit{et al.} in \cite{boehmermod, boehmerfq}.

\subsection{Standard method and Böhmer's method}\label{bohmer}
The standard method consists of normalizing the Friedmann equation as it appears in (\ref{norm1}), and from there choosing the variables \cite{bahamondedynamical, boehmermod}; this method is attributed to Amendola \textit{et al.} in \cite{amendoladynamic}. However, we must take into account that the energy density $\rho$ is the total of all fluids in the universe, that is, $\rho=\sum_{i}\rho_{i}$; with $i \in (1,2,...)$, where radiation and matter densities are tipically considered. Therefore, if the number of fluids is $N$, we will have $N$ dynamical variables arising from this term alone. From the geometric term of the model, which would correspond to the energy density of the modified gravitational field; one or two dynamical variables may arise. Thus, if we consider the case with two fluids (matter and radiation) along with the contribution from the model, we would have the following as possible variables:
\begin{align}
x=\frac{8\pi \rho_{m}}{3H^2}, \quad y=\frac{8\pi \rho_{r}}{3H^2}, \quad Z=\frac{F(Q)}{6H^2}, \quad W=-\frac{QF'(Q)}{3H^2},
\end{align}
here we have chosen the notation $Z$ to distinguish this variable from the redshift $z$. Note that $Z+W=\Omega_{\rm{DE}}$, reducing to $\Omega_{\Lambda}$ in the standard model. Introducing these variables, equation \ref{norm1} takes the form:
\begin{align}
x+y+Z+W=1; \quad x, y\geq 0. \quad 1-Z-W\geq 0,
\end{align}
where the last constraint is due to the positivity of the variables $(x,y)$, and it ultimately constraints parameters in the functional $F(Q)$. That is, if $F'(Q)<0$ ($W>0$), and if $F'(Q)>0$ ($W<0$). similarly $Z<0$ if $F(Q)<0; \quad \forall Q$, and $Z>0$ if $F(Q)>0 \quad \forall Q$.

In the case of the proposed exponential model, previously worked out in \cite{vasquezoliv, OliverosAcero}, the functional $F(Q)$ is given by:
\begin{align}
F(Q)=2\Lambda e^{-(b\Lambda/Q)^{n}}
 \end{align}
where $b,n$ are real numbers. Analytical solutions have been found in cases $n=1, |b|<1$ and $n=2, |b|<1$ \cite{vasquezoliv, OliverosAcero}, and a general study of the background cosmological parameters and linear matter perturbations is performed using the model. Generally, $F'(Q)=n(b\Lambda)^{n}Q^{-n-1}F(Q)\geq 0$ if $n=2k$ (even), regardless of the sign of $b$. However, if $n=2k+1$ (odd), then with $b<0$, $F'(Q)\leq 0$, making $W\geq 0$. For this case, the expression $Z+W$ satisfies the condition $1\geq Z+W\geq 0$. The variable $W$ is equivalent to $W=-2F'(Q)$, since $Q=6H^2$. Likewise, for our exponential model we have $F'(Q)=n(b\Lambda)^{n}Q^{-n-1}F(Q)=n(b\Lambda)^{n}Q^{-n}Z$, and therefore the relation holds: $W=-2m(b;\Lambda;n,Q)Z$. The function $m(b;\Lambda;n,Q)$ depends on the model parameters; for example, for $n=1$ we have $W=-2b\Lambda Z/6H^2$. The question of whether this system can be reduced to an autonomous system is valid. We observe that $W$ depends on $Z$ but also on the Hubble parameter, and we cannot eliminate this dependence, which results in the non-autonomy of the system. To remedy this, we can use a strategy proposed in \cite{boehmermod,boehmerfq}.
Using this strategy, we define the following dynamical variables:
\begin{align}
x=\frac{8\pi \rho_{m}}{3H^2}, \quad y=\frac{8\pi \rho_{r}}{3H^2}, \quad Z=\frac{F(Q)}{6H^2}, \quad m=\frac{H^2}{\Lambda+H^2}.
\end{align}
The variable $m$ is such that when $H^2 \gg \Lambda$, it satisfies the limit $m\rightarrow 1$. In the present, we have $m_{0}=\frac{1}{3\Omega_{\Lambda,0}+1}$, and at the de Sitter limit, using the expression for $H^2$ previously obtained in \cite{vasquezoliv}, we find $m_{dS}=\frac{1}{3\Omega_{\Lambda,0}\left(1-\frac{3}{2}b-\frac{13}{8}b^2\right)+1}$, with  $b<1$ and $n=1$ for the model (for $\Lambda$CDM, we have $m_{DE} \rightarrow 1/4$). Rewriting $H^2=m\Lambda/(1-m)$ and since $Z(H^2)$ via the function $F(Q)$, the system is thus reducible by writing $Z$ in terms of $m$:
\begin{align}
    Z=\frac{(1-m)}{3m}e^{-[b(1-m)/6m]^{n}}
\end{align}
In this way, we have removed the variable $W$, since the derivative $F'(Q)$ is proportional to $Z$ in the exponential model. For $n=1$, the constraint equation then becomes:
\begin{align}\label{constraintfr}
    x+y+\left(1-\frac{b(1-m)}{3m}\right)\frac{(1-m)}{3m}e^{-b(1-m)/6m}=1.
\end{align}
Now the system is described by the three dynamical variables $\{x, y, m\}$, and since we can cast $y=1-x-\left(1-\frac{b(1-m)}{3m}\right)\frac{(1-m)}{3m}e^{-b(1-m)/6m}$, the dynamical equation w.r.t to the customary evolution parameter $\eta=\ln{a}$ for $y$, can be written in terms of the pair $\{x,m\}$:
\begin{align}
    \frac{dy}{d\eta}=-\frac{dx}{d\eta}-\frac{du(m;b)}{d\eta}
\end{align}
where we have introduced for simplicity the function,
\begin{align}
    u(m;b)=\left(1-\frac{b(1-m)}{3m}\right)\frac{(1-m)}{3m}e^{-b(1-m)/6m},
\end{align}
which is parametrized by $b$. Using the chain rule we identify that the dynamics of variable $y$ is encoded in the dynamics of the pair $\{x,m\}$ as follows:
\begin{align}
    \frac{dy}{d\eta}=-\frac{dx}{d\eta}-\frac{du(m;b)}{dm}\frac{dm}{d\eta}.
\end{align}
Note that by definition, $\Omega_{DE}(m;b)=\left(1-\frac{b(1-m)}{3m}\right)\frac{(1-m)}{3m}e^{-b(1-m)/6m}$, and the positivity of the variables $(x,y)$ results in the upper bound, $\Omega_{DE}\leq 1$; which is actually a bound or constraint for the parameter $b$. The resulting dynamical equations for the variables are:
\begin{align}
\frac{dx}{d\eta}=-3x-2x\frac{\dot{H}}{H^2}, \\
\frac{dm}{d\eta}=2m(1-m)\frac{\dot{H}}{H^2}.
\end{align}
Here $\eta=\ln{a}$ is the customary evolution parameter. The term $\frac{\dot{H}}{H^2}$ is obtained by working out the expression $2QF''(Q)+1+F'(Q)$ in second Friedmann equation \ref{friedF}, in terms of $Z$. Therefore, given:
\begin{align}\label{fprime}
    F'(Q)=n(b\Lambda)^{n}Q^{-n}Z,
\end{align}
We take a further derivative with respect to Q by inserting back $Z=F(Q)/Q$ and proceed as,
\begin{align}
       F''(Q)=-(n+1)n(b\Lambda)^{n}Q^{-n-2}F(Q)+n(b\Lambda)^{n}Q^{-n-1}F'(Q),\\
    \Rightarrow QF''(Q)=-(n+1)n(b\Lambda)^{n}Q^{-n-1}F(Q)+n(b\Lambda)^{n}Q^{-n}F'(Q).
\end{align}
Inserting back (\ref{fprime}), we get:
\begin{align}
QF''(Q)=-n(n+1)(b\Lambda)^{n}Q^{-n-1}F(Q)+n^2(b\Lambda)^{2n}Q^{-2n}F(Q).
\end{align}
Grouping everything we leave the following term,
\begin{align}
    \label{fprime2}
    2QF''(Q)=2(b\Lambda)^n[-n(n+1)Q^{-n-1}+n^2(b\Lambda)^{n}Q^{-2n}]F(Q).
\end{align}
Particularly for $n=1$, using expressions (\ref{fprime}, \ref{fprime2}) we insert them back into their respective terms and find:
\begin{align}\label{FQQ}
2QF''(Q)+1+F'(Q)=-2b\Lambda Q^{-1}Z+2b^2\Lambda^2Q^{-2}Z+1,
\end{align}
as casted in terms of $\{x,m\}$ gives:
\begin{align}
2QF''(Q)+1+F'(Q)=\frac{b(1-m)^2e^{-b(1-m)/6m}[b(1-m)-6m]+54m^3}{54m^3}.
\end{align}
Therefore we get,
\begin{align}\label{hdoth}
\frac{\dot{H}}{H^2}=\frac{54m^3[x/2+2u(m)-2]}{t(m)}; 
\end{align}
where $t(m)=e^{-b(1-m)/6m}[b^2(1-m)^3-6b m(1-m)^2]+54m^3$. Putting everything together, the autonomous system is comprised by the following system of equations, completely written in terms of the pair $\{x,m\}$:
\begin{align}\label{systemiv}
\frac{dx}{d\eta}=-x\frac{[3t(m)+54m^3x+216m^3(u(m)-1)]}{t(m)}, \\
\frac{dm}{d\eta}=\frac{108m^4(1-m)(x/2+2u(m)-2)}{t(m)}.
\end{align}
The phase space of the system can be explored by considering conditions on the dynamical variables. In the table \ref{phasespace1}, values of $x$ and $y$ result in conditions on variable $m$ via the function $u(m)$. 
\begin{table}[h!]
\centering
\begin{tabular}{|c|c|c||}
\hline
$x=\Omega_{m}$ & $y=\Omega_{r}$ & Condition on $m$ \\
\hline
0 & 1 & $\left(1-\frac{b(1-m)}{3m}\right)\frac{(1-m)}{3m}e^{-b(1-m)/6m}=0$ \\
\hline
1 & 0 & $\left(1-\frac{b(1-m)}{3m}\right)\frac{(1-m)}{3m}e^{-b(1-m)/6m}=0$ \\
\hline
0 & 0 & $\left(1-\frac{b(1-m)}{3m}\right)\frac{(1-m)}{3m}e^{-b(1-m)/6m}=1$ \\
\hline
\end{tabular}
\caption{Conditions on $m$ by considering values of dynamical variables $\{x,y\}$}
\label{phasespace1}
\end{table}

\begin{figure}[h!]
\centering
\includegraphics[scale=0.7]{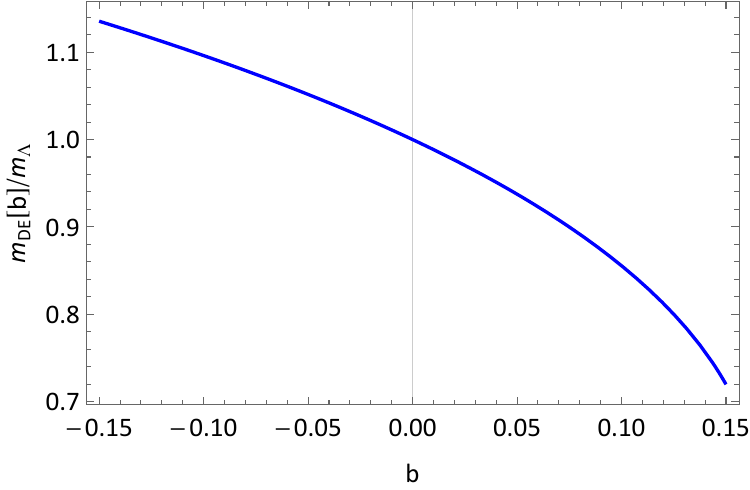}
\caption{Values for the de Sitter point of the dynamical system as a function of $b$. The function $m_{DE}$ goes through the point $1/4$ when $b=0$.}
\label{desitterpoint}
\end{figure}

For the first two conditions on $m$, they share the same solutions degenerately. Since $e^{-b(1-m)/6m}\rightarrow 0$ only for $m=0$, making $\Omega_{DE}\rightarrow \infty$ is not a possible solution. Another solution is $m^{*}=\frac{2b}{3+2b}$, which is not an equilibrium/fixed point of the system, so we also discard it as a solution of interest. The solution $m=1$, while the simplest, is a fixed point of the system. For the last condition, $m$ cannot be obtained straightforwardly given that the equation to solve is transcendental. We can solve it employing approximations in order to obtain the parametrized dependence $m(b)$, which for $\Lambda$CDM we have $m_{\Lambda}=1/4$. We thus approximate the exponential for $b\ll 1$ up to second order, and we obtain a third-order algebraic equation for $m$ of the form,
\begin{align}
\frac{m^2}{3}(1-m)-\frac{mb}{6}(1-m)^2+\frac{5b^2(1-m)^3}{216}\approx m^3.
\end{align}
Further approximations can be obtained since $|b|<1$, allowing us to neglect terms such as $b^2(1-m)^3$. The result is a second-order equation for $m$ whose solutions are dubbed $m_{\rm{DE}\pm}$ and given by the expression,
\begin{align}
m_{\rm{DE}\pm}=\frac{(1+b)\pm\sqrt{1-6b}}{b+8}.
\end{align}
And we choose $m_{\rm{DE}+}$ in correspondence to $\Lambda$CDM value, $m_{\Lambda}=1/4$, Likewise, for $b>1/6$ the point may result non-hyperbolic when analyzing stability. By plotting $m_{\rm{DE}+}/m_{\Lambda}$ for positive and negative values of $b$ we observe that the critical point $m_{\rm{DE}+}$ occurs below $m_{\Lambda}$ for $b>0$ and conversely for $b<0$ (see figure \ref{desitterpoint}). 

From the previous figure, we can visualize the effect of parameter $b$ on the model as studied before in \cite{OliverosAcero, vasquezoliv}. For positive $b$ in $n=1$ the model behaves as a quintessence field and for negative $b$, as a phantom field. This behavior also shows itself in the appearance of the de Sitter point for the dynamical system, as the point moves to values lower than $m_{\rm{DE}}$, leading to an effect on the cosmic expansion rate.

\section{Phase-space analysis and equilibrium points}\label{phasespace}

The system (\ref{systemiv}) exhibits equilibrium points when the condition, $\dot{\textbf{x}}(t)=0$ with $\{x, m\}\in \textbf{x}$ is met. From here, the values $x=0, m=1, m=m_{\rm{DE}}$ occur to form equilibrium points. Allowing us to list them all:
\begin{itemize}
    \item \textbf{Radiation dominated point:} this point, $R$ occurs at the coordinates $\{0,1\}$ of phase space. This point makes $d_{\eta}x=d_{\eta}m=0$.

    \item \textbf{Matter dominatied point:} called $M$, it occurs at the coordinates $\{1,1\}$ of phase space. Likewise it makes $d_{\eta}x=d_{\eta}m=0$.

    \item \textbf{de Sitter point:} this point, $P_{dS}$ occurs for the coordinates $\{0, m_{\rm{DE}}\}$, where the value $m_{\rm{DE}}$ results as a solution to the transcendental equation $u(m)=1$. This solution depends on parameter $b$ as shown in figure \ref{desitterpoint}, therefore, the contributions of the model appear as expected, at late times.
\end{itemize}

\begin{figure}[h!]
\centering
\includegraphics[scale=0.58]{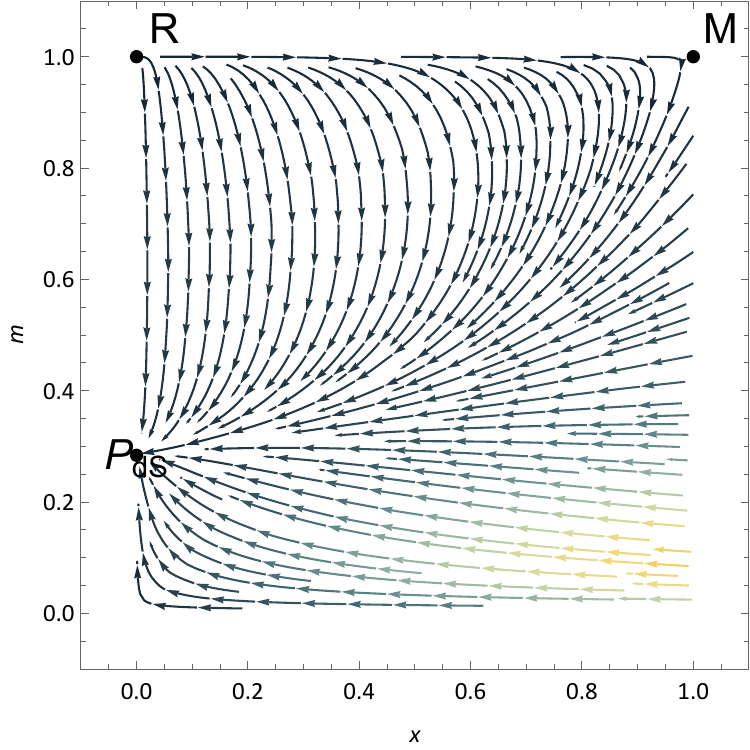}
\hspace*{0.1 cm}
\includegraphics[scale=0.58]{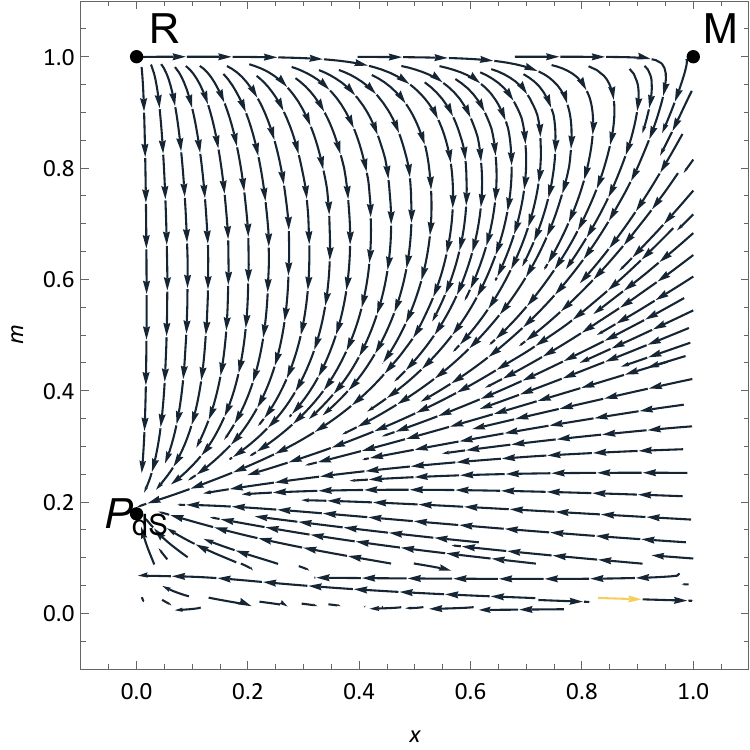}
\caption{Phase-space portraits of the dynamical system with contributions from the model. For negative values of $b$ [Left]. For positive $b$ values [Right].}
\label{phasespaceport}
\end{figure}

\begin{figure}[h!]
\centering
\includegraphics[scale=0.58]{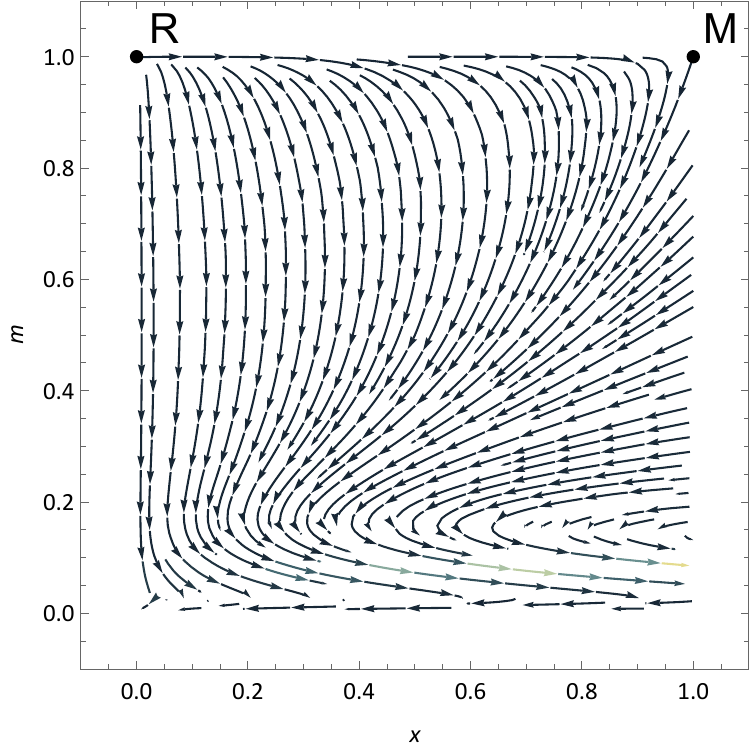} 
\caption{Phase space for $b=1/3$. The de Sitter attractor point disappears, since variable $m$ becomes complex and the point is non-hyperbolic.}
\label{phasespacnoDS}
\end{figure}

Figure \ref{phasespaceport} allows us to observe the equilibrium points in the phase space portraits of the system and how the de Sitter point changes as parameter $b$ varies. We have used values $b=+0.1, -0.1$ for each portrait, and in both cases a heteroclinic orbit is obtained, driving the universe through the points $R\rightarrow M\rightarrow P_{dS}$. In addition, the influence of the parameter $b$ is shown, as previously discussed and presumably for $b>1/6$ the de Sitter point may disappear from the dynamical system since $m_{\rm{DE}+}$ becomes complex. Clearly, the solution $m_{\rm{DE}}$ is limited only to describe real roots as long as $b<1/6$, for $b>1/6$ the point $P_{dS}$ vanishes from the phase space portrait. On the other hand, for illustrational purposes, if values $b\gg1$ occur (or at least $b=1$), it is not guaranteed that a real solution for $u(m;b)=1$ exists; meaning that there would no equilibrium point by itself (See figure \ref{phasespacnoDS}). In this way, equilibrium points $\{R,M,P_{dS}\}$ form an invariant set $S\in \mathbb{R}^{2}$ where the solutions are bounded by the heteroclinic orbit joining the points. Therefore, the model must qualitatively correspond to the same dynamics as in $\Lambda$CDM with variations at late times, as we observe in the phase space portraits.

In order to analyze the stability of the critical points, we write the following Jacobian matrix,

\begin{align}
J_{b}=
\begin{pmatrix}
\partial_{x}f_{1} & \partial_{m}f_{1}\\
\partial_{x}f_{2} & \partial_{m}f_{2},
\end{pmatrix}
\end{align}
with $f_{1}=-x\frac{[3t(m)+54m^3x+216m^3u(m)-216m^3]}{t(m)}$ and $f_{2}=\frac{108m^4(1-m)(x/2+2u(m)-2)}{t(m)}$. Thus, we evaluate the Jacobian at each equilibrium point such that\footnote{The derivatives and their evaluation were realized using Mathematica for convenience, since the expresions turn out to be very long.}
\begin{align}\label{jacobians}
J|_{R}=\begin{pmatrix}
1 & 0\\
0 & 4
\end{pmatrix}, \quad J|_{M}=\begin{pmatrix}
-1 & 4/3\\
0 & 3
\end{pmatrix}, \quad J|_{P_{dS}}=\begin{pmatrix}
g_{1}(b) & 0\\
g_{3}(b) & g_{4}(b)
\end{pmatrix}.
\end{align}
where $g_1(b)$, $g_3(b)$ and $g_4(b)$ are given by
\begin{equation}
\begin{aligned}
g_1(b)=&-\dfrac{3\Big[(-7+A)^{3} b^{2}+14(-7+A)^{2} b\, C-24(8+b) C^{2}
+6 C^{3}(4+3B)\Big]}{(-7+A)^{2} b \big[6(1+A)+(-1+A)b\big]-54 C^{3} B},
\end{aligned}
\end{equation}

\begin{equation}
g_3(b)=\frac{54\,(7-A)\,C^{4}}{(8+b)^{2}\!\left[54\,C^{3}-(-7+A)^{2} b\big(6(1+A)+(-1+A)b\big)\,B\right]}
\end{equation}

\begin{equation}
\begin{aligned}
g_4(b)=& -\frac{12\,C}{(8+b)\big[(-7+A)^2 b(6(1+A)+(-1+A)b)-54 C^{3}B\big]^2}\times\\[4pt]
&\quad\Big\{-6(-7+A)^3 b^2 C\big[(-7+A)(-34+6A+b)+6(-23+A-2b)C B\big] \\
&\qquad+18(-7+A)^2 b C^{2}\big[4(-7+A)^2-27(8+b)C B\big]\\
&\qquad+(-7+A)^4 b^3\big[4(-7+A)^2-3(8+b)C B\big]\\
&\qquad+324\,C^{5}\big[-3(-6+2A+b)F-2(-7+A)G\big]
\Big\},
\end{aligned}
\end{equation}
with
\begin{equation}
A=\sqrt{1-6b},\qquad B=e^{\tfrac{1-A}{6}},\qquad
C=1+A+b,\qquad F=e^{\tfrac{1-A+b}{3}},\qquad G=e^{\tfrac{1-A+2b}{6}} 
\end{equation}

\begin{figure}[h!]
\centering
\includegraphics[scale=0.58]{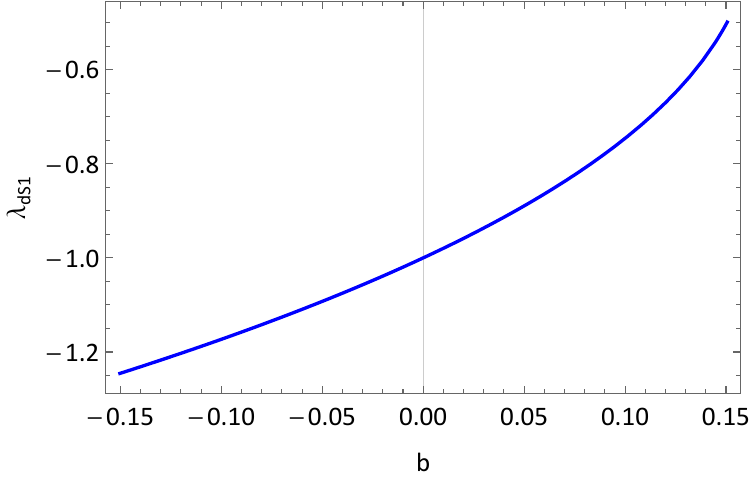}
\hspace*{0.1 cm}
\includegraphics[scale=0.58]{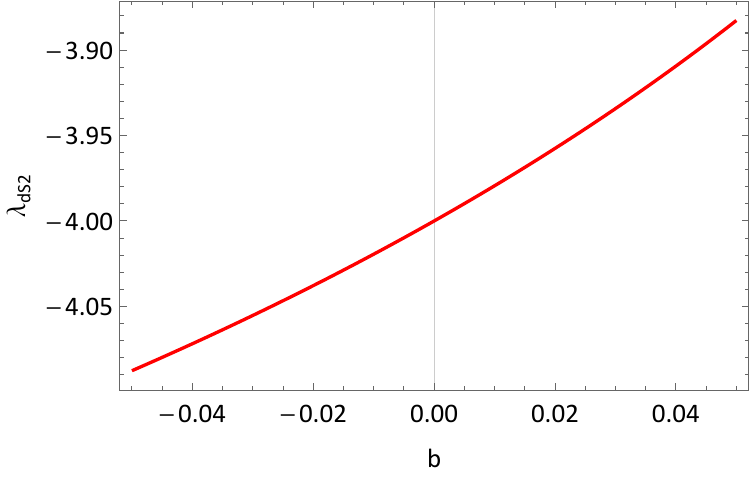}
\caption{Eigenvalues of the Jacobian $J|_{P_{dS}}$ for negative and positive values of  $b$.}
\label{lambdapds}
\end{figure}

The next step is to obtain the eigenvalues of the Jacobian, for which in the case of de Sitter point $P_{dS}$ we must evaluate the resultant functions of parameter $b$. In general we obtain, 
\begin{align}
(\lambda_{R}-1)(\lambda_{R}-4)=0, \quad (\lambda_{M}+1)(\lambda_{M}-3)=0, \quad (g_{1}(b)-\lambda_{dS})(g_{4}(b)-\lambda_{dS})=0.
\end{align}
Therefore we have: for $R$ the eigenvalues are $\{1,4\}$ thus the point is \textbf{unstable} or a \textbf{repeller}. For $M$, we got $\{-1,3\}$, a saddle point, serving as a transition between matter dominated epochs to late times when dark energy dominates. The eigenvalues for the matrix evaluated at $P_{dS}$ are given by the functions $\{g_{1}, g_{4}\}$, and to determine their signs we plot both functions (see \ref{lambdapds}). The plot shows that for both $b>0$ and $b<0$ the eigenvalues are negative, indicating that $P_{dS}$ is a \textbf{stable point} or an \textbf{attractor} in the future; independent of the values for $b$.

\subsection{Cosmological parameters of the system}

\begin{figure}[h!]
\centering
\includegraphics[scale=0.58]{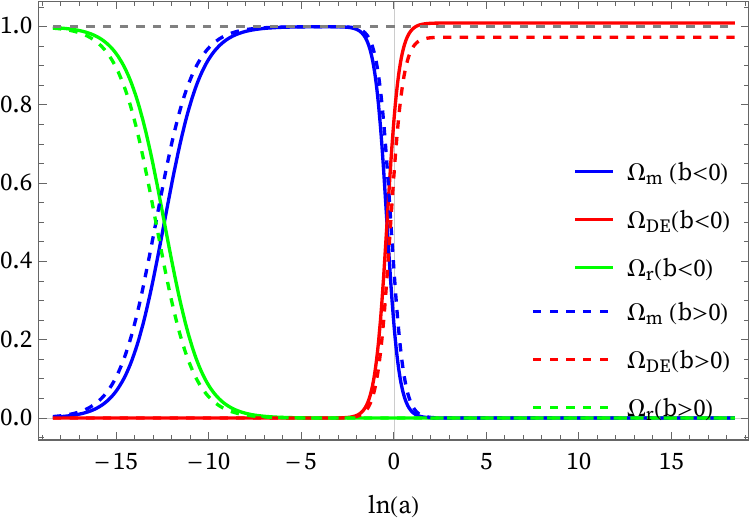} 
\caption{Evolution of density parameters for matter and radiation, additionally the geometrical contribution from the model. $b>0$ [Dashed lines] and $b<0$ [Solid lines]}
\label{omegall}
\end{figure}

Usually, parameters such as the effective state parameter $w_{\rm{eff}}$ and deceleration $q$ are evaluated at each equilibrium point to determine the state of the universe at the given point. Using the following expressions for $w_{\rm{eff}}$ and $q$,
\begin{align}
q=-1-\frac{\dot{H}}{H^2}, \quad w_{\rm{eff}}=-1-\frac{2\dot{H}}{3H^2},
\end{align}
then we can rewrite them in terms of the dynamical variables $\{x,m\}$. Leaving us with the expressions,
\begin{align}
q(x,m)=-\frac{(54m^3[x/2+2u(m)-2]+t(m))}{t(m)}, \\
w_{\rm{eff}}(x,m)=-\frac{(108m^3[x/2+2u(m)-2]+3t(m))}{3t(m)}.
\end{align}
With respect to unstable point $R$ we have, $t(m)|_{R}\rightarrow 54, u(m)|_{R}\rightarrow 0$; therefore $q_{R}(0,1)=1$, and $w_{\rm{eff}}(0,1)=1/3$. For the saddle point $M$, $t(m)|_{M}\rightarrow 54, u(m)|_{M}=0$ and $x=1$; we obtain $q_{M}(1,1)=1/2$ y $w_{\rm{eff}}(0,1)=0$. Finally at the attractor point $P_{dS}$, we have that $u(m)|_{P_{dS}}=1, x=0$; such that the deceleration parameter is $q_{P_{dS}}=-1$ and the effective state parameter, $w_{\rm{eff}}(0, m_{DE})=-1$. With this analysis we confirm that the model correctly reproduces the main dominance epochs known in the universe. As we want to fully analyze the evolution, allow us write the dynamical equations for the variables $\{x,y,m\}$, by reintroducing dynamical variable $y$ pertaining to the radiation density parameter:
\begin{align}\label{xym}
\frac{dx}{d\eta}=-x\frac{[3t(m)-108m^3(3x/2+2y)]}{t(m)}, \\
\label{yevol}
\frac{dy}{d\eta}=-y\frac{[4t(m)-108m^3(3x/2+2y)]}{t(m)}, \\
\frac{dm}{d\eta}=-\frac{108m^4(1-m)(3x/2+2y)}{t(m)}.
\end{align}
We obtain the set (\ref{xym}) by plugging the constraint condition (\ref{constraintfr}) back into (\ref{hdoth}) and (\ref{systemiv}). This is,
\begin{align}
    \frac{x}{2}+2u(m)-2=\frac{x}{2}-2x-2y=-\left(\frac{3x}{2}+2y\right).
\end{align}
In the particular case of equation (\ref{yevol}) we use variable $y=8\pi\rho_{r}/3H^2$, and then by taking the derivative with respect to $d\eta=Hdt$ we obtain,
\begin{align}
\frac{dy}{d\eta}=-4y-2y\frac{\dot{H}}{H^2},
\end{align}
where the first term comes from the continuity equation $\dot{\rho}_{r}=-4H\rho_{r}$. These equations are solved numerically with initial conditions:  $y(0)=10^{-6}$, $x(0)=1-\Omega_{DE}(b,0)-y(0), m(0)=m_{0}$. The combined result is shown in the plot (\ref{omegall}), where we have used the values $b=\pm 0.1$ to show that; for $b>0$ it reaches the limit of $+1$, while for $b<0$ we obtain a behavior such that $\Omega_{DE}\gtrsim 1$ as previously discussed.

\begin{figure}[h!]
\centering
\includegraphics[scale=0.6]{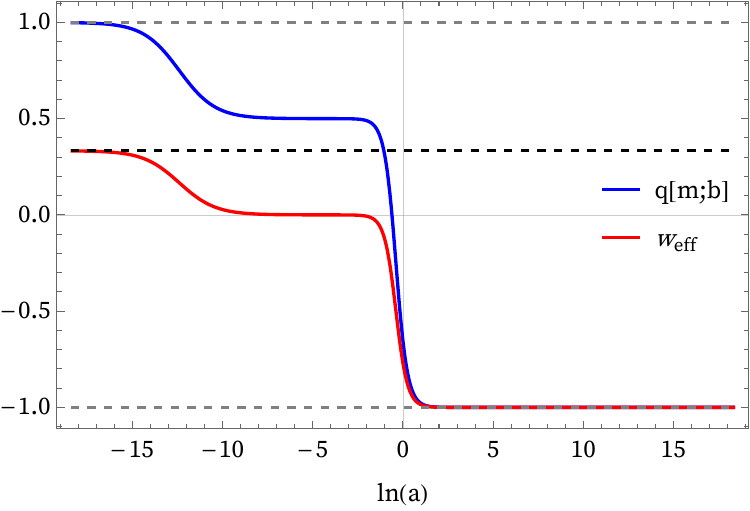}  
\caption{Evolution of parameters $q$ and $w_{eff}$. (Values $+1, -1, 1/3$ are shown as a reference.)}
\label{qweff}
\end{figure}

In the case of cosmological parameters $(q, w_{\rm{eff}})$, the evolution is shown in figures (\ref{qweff}). Here, each distinct epoch is clearly marked using dotted lines as a reference. We have used $b=-0.1$, however, changing for positive values the behavior does not change in terms of limiting values. At this level, hence, the model shows the same features of $\Lambda$CDM and represents a viable model, since at perturbative level it also shows slight deviations as studied in \cite{vasquezoliv}. Both parameters are related through the term $\dot{H}/H^2$, so both have the minimum value of $-1$ under the built model; therefore, even in the infinite future the null energy condition holds ($w_{\rm{eff}}\geq -1$). 

\section{Exponential $f(Q)$ model with DM-DE linear interactions}\label{dmdeinterac}

Although successfully supported by observations, the $\Lambda$CDM still has a number of unsolved issues. A classic example is the \textit{cosmological constant problem} (see \cite{copeland, Weinberg1972} and references therein). This problem generally led the community to propose cosmological models with a variable dark energy, as it is the case of quintessence \cite{zlatev}. However, introducing temporal variations over dark energy, suggests the following problem: 

\begin{quote}
If observations show that $\Omega_{m,0}\sim 0.3$, while $\Omega_{\Lambda,0}\sim 0.7$ at the present; with both components evolving at different rates. \textit{How is it possible that they now coincide after billions of years of evolution?}
\end{quote}

This apparent coincidence suggests that their values had to be somehow \textit{adjusted} during the early stages of cosmological evolution \cite{Wang2016}. This problem motivated the introduction of models with interactions between dark matter and dark energy, with the purpose of explaining their apparent relationship in relative densities. It is then supposed, that an interaction between these two components may result in a solution to the coincidence problem. In these model, generally dark matter density $\rho_{\rm{dm}}$ and dark energy $\rho_{\rm{DE}}$ (in our case, of geometric origin), obey the following continuity equations:
\begin{align}
\dot{\rho}_{c}+3H\rho_{c}=\mathcal{Q}, \quad \dot{\rho}_{DE}+3H(1+w_{DE})\rho_{DE}=-\mathcal{Q}.
\end{align}
Here,  $\mathcal{Q}$ is the interaction kernel; it represents a function that determines the interaction. Note that the sum $\rho_{\rm{dm}}+\rho_{\rm{DE}}$ follows the continuity equation $\dot{\rho}+3H(\rho_{\rm{dm}}+(1+w_{\rm{DE}})\rho_{\rm{DE}})=0$. The precise form of the interaction kernel $\mathcal{Q}$ has not been obtained from first principles, mainly because the nature of dark energy and dark matter is unknown. Therefore, the kernel rests upon phenomenological constructions that may, at least, respect some expected features. For example, observations indicate that the $\Lambda$CDM model is the best-fitted model, and one for which dark energy takes the form of the cosmological constant. This means that if the interaction does exist, it must be small enough and slowly evolving, in order to coincide with observational data \cite{Wang2016}. Between the commonly used kernels, they have a general form as follows (see \cite{Wang2024}, and references therein):
\begin{align}
\mathcal{Q}:=\mathcal{Q}(\rho_{c},\rho_{DE};\xi_{i}); \quad \xi_{i}\neq 0 \wedge \xi_{i}\in \mathbb{R}.
\end{align}
That is, a function of the DM-DE densities; usually ordinary functions that could be linear or non-linear. There is a wide selection of kernels, with distinct features that are rigourously analyzed in the series of papers \cite{vander, vander2, vander3}. Also included, there is a coupling term that parametrizes the density function ($\xi_{i}$), this term indicates the intensity of the interaction. It could be only one term such that $\xi_{i}=\xi$ or several, depending on the model. Initially, the motivation for introducing DM-DE interactions results from the coincidence problem, however, it has been recently studied in the context of \textit{Hubble tension} and $S_{8}$ tension  \cite{Wang2024}. Likewise, the interaction seems possible according to recent data from DESI  \cite{Sabogal2}. 

\subsection{Linear interaction kernel}
In order to include interaction in the system, we first consider that matter density contains both baryons and dark matter contributions. such that:
\begin{align}
\rho_{m}=\rho_{b}+\rho_{c}, \\
\Rightarrow \dot{\rho}_{m}=\dot{\rho}_{b}+\dot{\rho}_{c}=-3H\rho_{b}-3H\rho_{c}+\mathcal{Q}, \\
\label{dmevol}
\Rightarrow \dot{\rho}_{m}=-3H\rho_{m}+\mathcal{Q}.
\end{align}
Here, $\rho_{m}$ is the total matter density and $\rho_{b}$ is the baryon contribution as well as $\rho_{c}$ is the cold-dark matter contribution.  In this way, the differential equation for the dynamical variable $x$ transforms into,
\begin{align}
\frac{dx}{d\eta}=-3x-2x\frac{\dot{H}}{H^2}+\frac{8\pi\mathcal{Q}}{3H^3}.
\end{align}
In general terms, the interaction kernel is a function written in terms of dynamical variables $\mathcal{Q}(m,x)$. Likewise, when $\mathcal{Q}>0$, the DM sector absorbs energy from the DE sector and the opposite occurs for $\mathcal{Q}<0$. Many phenomenological forms have been proposed, for example, the kernel $\mathcal{Q}_{1}=\xi H \rho_{c}$; which is dominant at early stages for $\xi>0$. The kernel $\mathcal{Q}_{2}=\xi H \rho_{\rm{DE}}$, is dominant at late times. Since our cosmological analysis focuses on late times, in a universe dominated by dark energy it is convenient to choose the second kernel $\mathcal{Q}_{2}$, indicating that the DM–DE interaction occurs at a variable rate in epochs close to the present, while in the past it tends to behave as a constant interaction given by the value $\rho_{\Lambda}$, according to the dynamics of the proposed model. An ambiguity arises when examining the strength of the interaction, since its sign determines whether the dark energy fluid dissipates energy and transfers it to the dark matter sector, or vice-versa, at a constant rate. Near the present epoch, depending also on the model parameter $b$, the interaction may \textit{reinforce} or \textit{weaken} the dominance of dark energy. Usually, the coupling term is taken as a constant; however, models have been proposed in which $\xi$ is a function of the redshift $z$, with the aim of alleviating the $H_{0}$ and $S_{8}$ tensions, making $\xi(z)$ change sign at the redshift where $\Omega_{m}/\Omega_{\rm{DE}}=1$ \cite{Sabogal2025}.

\subsubsection{Evolution of densities with interaction}With the interaction kernel $\mathcal{Q}=\xi H \rho_{\rm{DE}}$, the time evolution of the dark energy density is described by the following differential equation:
\begin{align}
(1+z)\rho'_{\rm{DE}}=(\xi+3[1+w_{\rm{DE}}])\rho_{\rm{DE}},
\end{align}
here the prime denotes a derivative with respect to redshift $z$. From this, we can solve for $w_{\rm{DE}}$, showing that the effect of the interaction is to modify the dark energy equation-of-state parameter, yielding an effective parameter given by:
\begin{align}
w^{(int)}_{\rm{DE}}=-1+(1+z)\frac{\rho'_{DE}}{3\rho_{DE}}-\frac{\xi}{3}.
\end{align}
The evolution of the equation-of-state parameter indicates that even in the $\Lambda$CDM limit, dark energy does not behave as a cosmological constant; however, the coupling $\xi$ must satisfy $|\xi|\ll 1$. We plot the equation-of-state parameter for $b=-0.1, +0.1$, and choose coupling values $\xi=-0.05, 0, -0.02$, noting that negative values of the coupling are favored by recent DESI collaboration data, according to the analysis in \cite{Sabogal2025}. In the case $b=-0.1$, which in the absence of interaction produces a phantom-like equation of state $w_{\rm{DE}}<-1$, the curve crosses the $-1$ barrier, making a transition toward a quintessence-like behavior. On the other hand, for $b=+0.1$, the behavior remains quintessence-like even in the presence of interaction. This quintom-like transition is also favored by recent DESI collaboration data using Baryon Acoustic Oscillations (BAOs) \cite{Sabogal2, nuo}.

\begin{figure}[h!]
\centering
\includegraphics[scale=0.58]{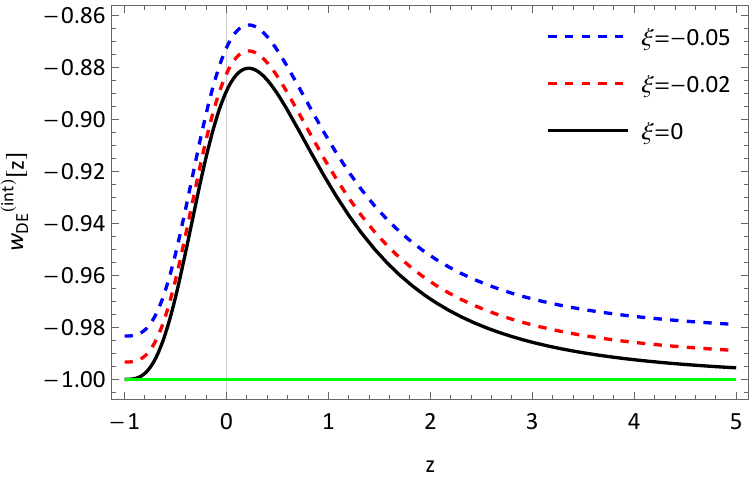} 
\hspace*{0.1 cm}
\includegraphics[scale=0.58]{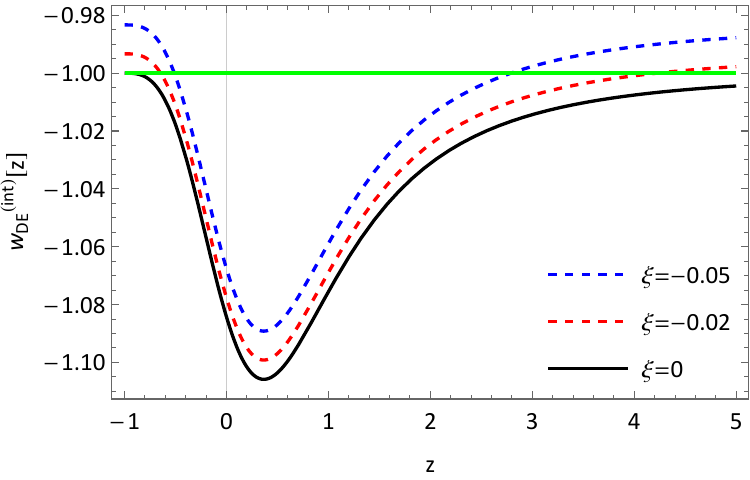}
\caption{Evolution of stateparameter $w_{\rm{DE}}$ for different values of the coupling $\xi$ ($b=0.1$ [Left] and $b=-0.1$ [Right.])}
\label{wDEint}
\end{figure}

\begin{figure}[h!]
\centering
\includegraphics[scale=0.80]{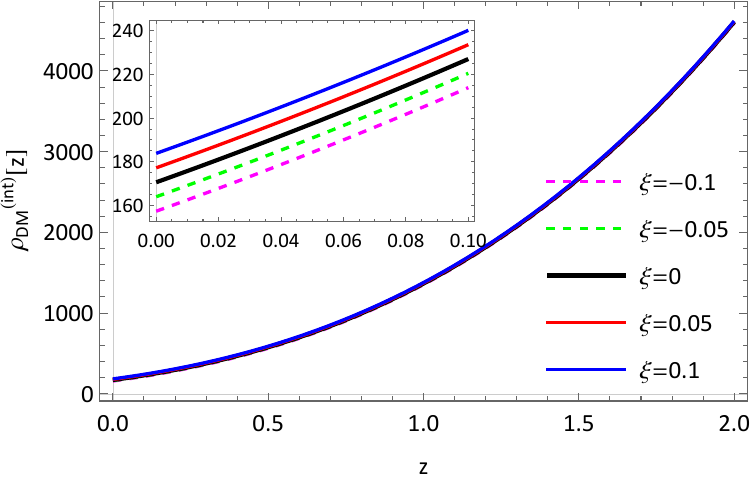} 
\caption{Evolution of dark matter density for different values of $\xi$.}
\label{RhoMatterInt}
\end{figure}

For the evolution of the dark matter component, we have the differential equation, obtained by plugging the interaction kernel $\mathcal{Q}$ into (\ref{dmevol}), and using the known transformation $d/dt\rightarrow -(1+z)H \, d/dz$:
\begin{align}
(1+z)\rho'_{c}-(3\rho_{c}-\xi\rho_{DE})=0,
\end{align}
where the derivative is taken with respect to the redshift $z$. We must also take into account that the baryonic matter component satisfies the continuity equation $(1+z)\rho'_{b}+3H\rho_{b}=0$. We solve the differential equation with the initial condition at $z=10$, given by $\rho_{m}(10)=\rho_{m,0}(1+10)^3$, noting that the effect of the coupling occurs only at late times. In Fig.~\ref{RhoMatterInt}, one can see that the influence of the interaction does not produce significant changes in the matter density; only at $z\ll 1$ it does occur that, for negative interaction parameters the density is lower than for positive interaction parameters. In this case we may interpret that for $\xi> 0$ the DM-component absorbs energy density from the DE-component while for $\xi < 0$ the opposite occurs.
\section{Phase-space analysis with DM-DE interactions}\label{interactingde}

\begin{figure}[h!]
\centering
\includegraphics[scale=0.58]{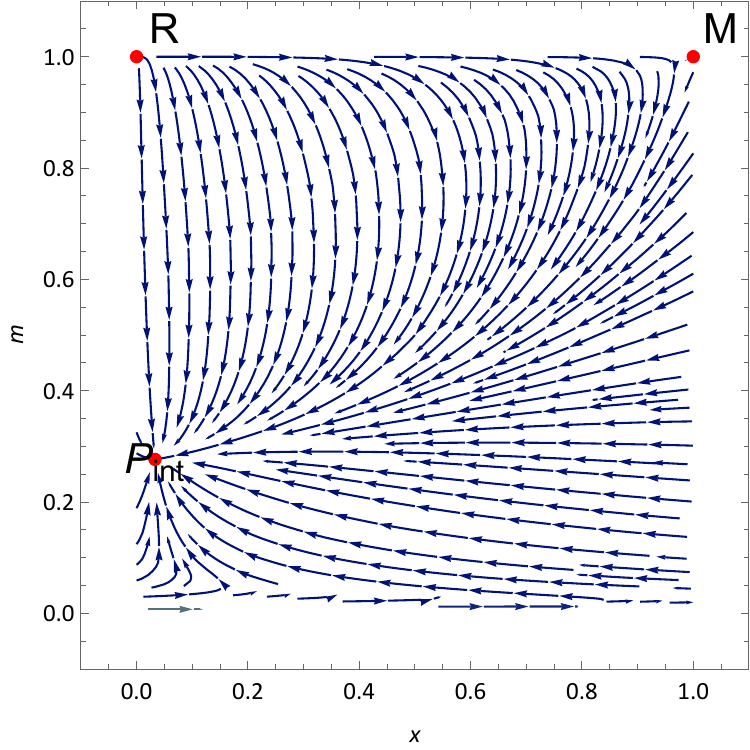} 
\hspace*{0.1 cm}
\includegraphics[scale=0.58]{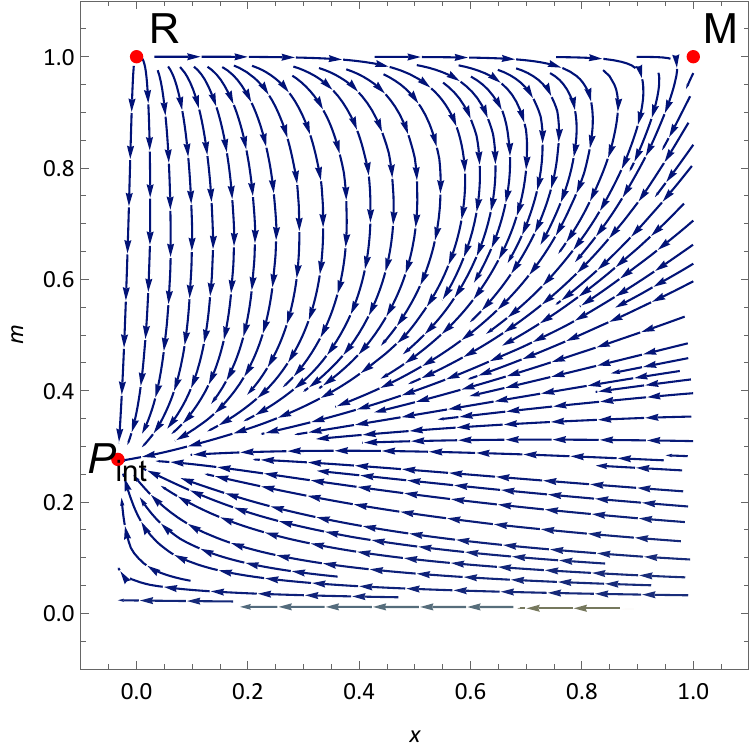}
\hspace*{0.1 cm}
\includegraphics[scale=0.58]{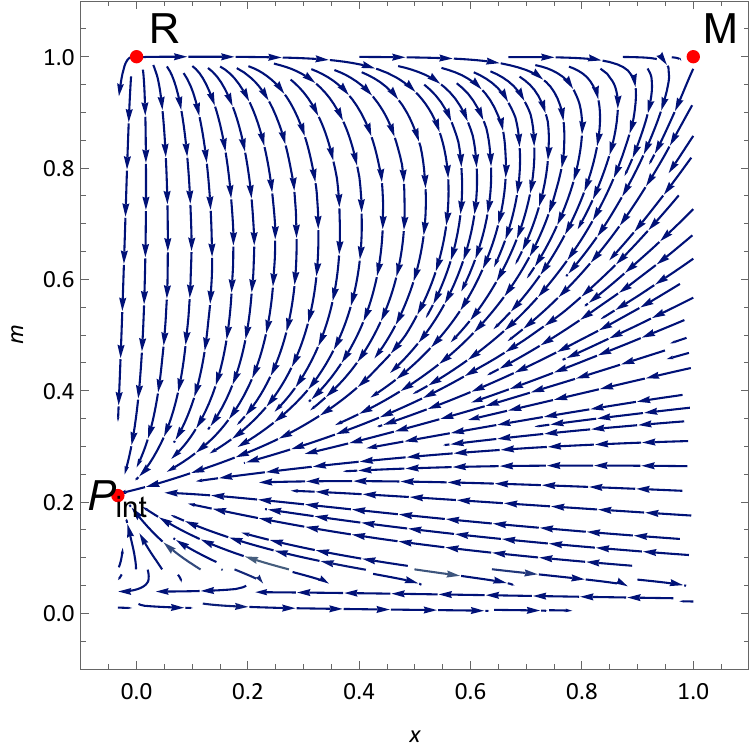} 
\hspace*{0.1 cm}
\includegraphics[scale=0.58]{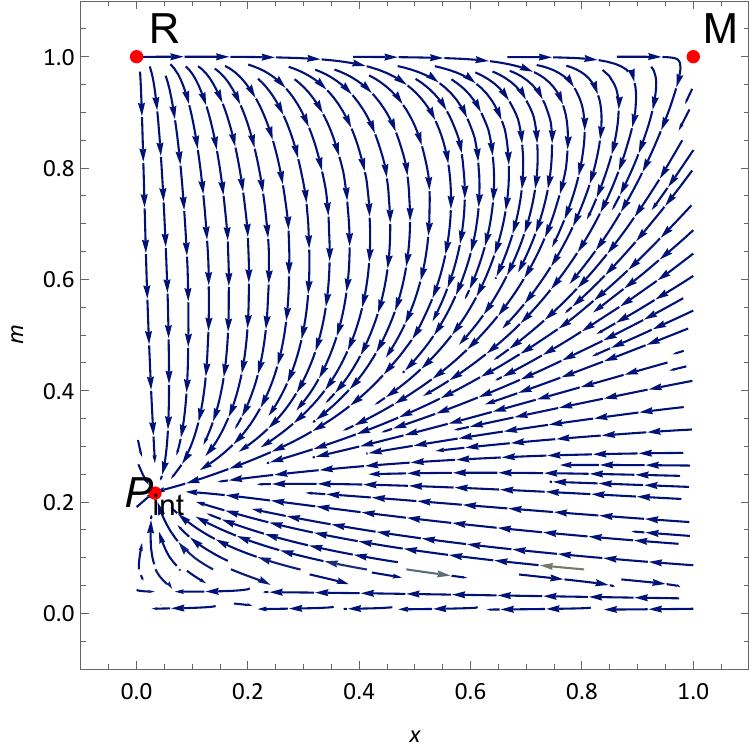}
\caption{Phase spaces of the dynamical system in the variables $(x,m)$ with linear interaction 
$\mathcal{Q}=\xi H\rho_{DE}$ for the following parameter combinations: 
(i) $b=-0.1$, $\xi=0.1$ [Left, Top], 
(ii) $b=-0.1$, $\xi=-0.1$ [Right, Top], 
(iii) $b=+0.1$, $\xi=-0.1$ [Left, Bottom], 
(iv) $b=+0.1$, $\xi=0.1$ [Right, Bottom].}
\label{phasespaceint}
\end{figure}

The interaction term transforms from $8\pi \mathcal{Q}/3H^3$, to $\xi\Omega_{\rm{DE}}$ by using the definition of the density parameter $\Omega_{\rm{DE}}=8\pi\rho_{\rm{DE}}/3H^2$. Hence, in terms of the dynamical variable $m$ we obtain the following dynamical equations:
\begin{align}
\frac{dx}{d\eta}=-x\frac{[3t(m)+108m^3(x/2+2u(m)-2)]}{t(m)}+\xi u(m), \\
\frac{dm}{d\eta}=\frac{108m^4(1-m)(x/2+2u(m)-2)}{t(m)}.
\end{align}
With the inclusion of the interaction, we note the following: (i) the radiation-dominated point $R$ remains in the system, (ii) the matter-dominated point $M$ with coordinates $(1,1)$ also remains unchanged, and (iii) the de Sitter point is modified by the interaction. Originally, the condition for the de Sitter point is $u(m)=1, x=y=0$, but this does not lead to an equilibrium point of the system in the presence of the interaction. Therefore, we must instead consider the condition $x/2=2-2u(m)$, which leads to
\begin{align}
    \frac{dx}{d\eta}=-3x+\xi u(m), \\
    \frac{dm}{d\eta}=0.
\end{align}
In order to obtain $d_{\eta}x=0$, it must hold that $3x=\xi u(m)$, and since $x=4-4u(m)$, we obtain the relationship $u(m)=1/(1+\xi/12)$ and consequently $x=4\xi/(12+\xi):=x_{int}$. Thus, the value of $m^{(int)}_{DE}$ is obtained by solving the algebraic equation,
\begin{align}
\frac{1-m}{3m}-\frac{(1-m)^2b}{6m^2}=\frac{1}{1+\xi/12}.
\end{align}
Multiplying by $6m^2$ and by $1+\xi/12$, and expanding $(1-m)^2=m^2-2m+1$, we obtain the form
\begin{align}
2m(1-m)(1+\xi/12)-(m^2-2m+1)b(1+\xi/12)=6m^2,
\end{align}
from which we see that the terms can be grouped into a part without the interaction $\xi$ and a part with it:
\begin{align}
-2m\left(1+\frac{\xi}{12}+b+\frac{\xi b }{12}\right)+b+\frac{b\xi}{12}+m^2\left(-4-b+\frac{\xi}{6}-\frac{\xi b}{12}\right)=0,
\end{align}
where the condition $b\xi\ll 1$ allows us to approximate the expression as
\begin{align}
m^2\left(-8-b-\frac{\xi}{6}\right)+2m\left(1+\frac{\xi}{12}+b\right)-b=0.
\end{align}
This leads to solutions in terms of the parameters $(b,\xi)$ of the form
\begin{align}
m_{\pm}=\frac{12+12b+\xi\mp \sqrt{144-864b+24\xi}}{96+12b+2\xi}.
\end{align}
We select the negative solution such that $m_{\pm}(b=0,\xi=0)=1/4$. We therefore see how the model and the interaction parameter influence the value of $m$, defined here as $m^{(int)}_{DE}$, referring to the interaction in contrast with the previous value. Finally, we obtain that the system has an equilibrium point $P_{int}:=\{x_{int}, m^{(int)}_{DE}\}$ in which $x_{int}\neq 0$ if $\xi\geq 0$, since for $\xi<0$ negative density parameters arise for dark matter. Hence, for $\xi\geq 0$ there is no complete dark energy dominance, because the interaction causes the energy density to be transferred to the dark matter. In Fig.~\ref{phasespaceint}, the phase portraits for different values of the parameters $(b,\xi)$ are shown. One can see how the coupling parameter $\xi>0$ shifts the equilibrium point toward positive values of $x$, while for $\xi<0$ the equilibrium point is shifted towards negative values of $x$, while positive and negative values of $b$ shift the point $P_{int}$ along the $m$ axis. If we strictly require the positivity of the parameter $\Omega_{m}$, then we must consider  only $\xi>0$ for this interaction model, so that the energy density is transferred to dark matter resulting in a decrease in dark energy density towards the future.

\begin{table}[h!]
\centering
\begin{tabular}{|c|c|c|c|c|c|}
\hline
\textbf{Critical points} & $x$ & $m$ & $w_{\rm{eff}}$ & $q$ & \textbf{Stability} \\
\hline
R & 0 & 1 & 1/3 & +1 & \textbf{Unstable} \\
\hline
M & 1 & 1 & 0 & 1/2 & \textbf{Saddle point} \\
\hline
$P_{dS}$ & 0 & $\approx \frac{(1+b)\pm\sqrt{1-6b}}{(b+8)} $ & -1 & -1 & \textbf{Stable} \\
\hline
\multicolumn{6}{|c|}{\textbf{With interaction}} \\
\hline
R & 0 & 1 & 1/3 & +1 & \textbf{Unstable} \\
\hline
M & 1 & 1 & 0 & +1/2 & \textbf{Saddle point} \\
\hline
$P_{int}$ & $4\xi/(12+\xi)$ & $\approx \frac{12+12b+\xi+\sqrt{144-864b+24\xi}}{96+12b+2\xi}$ & $<-1$ & $<-1$ & \textbf{Stable} \\
\hline
\end{tabular}
\caption{Summary of the critical points of the system without and with interaction.}
\label{tablesummarypoints}
\end{table}

\begin{figure}[h!]
\centering
\includegraphics[scale=0.59]{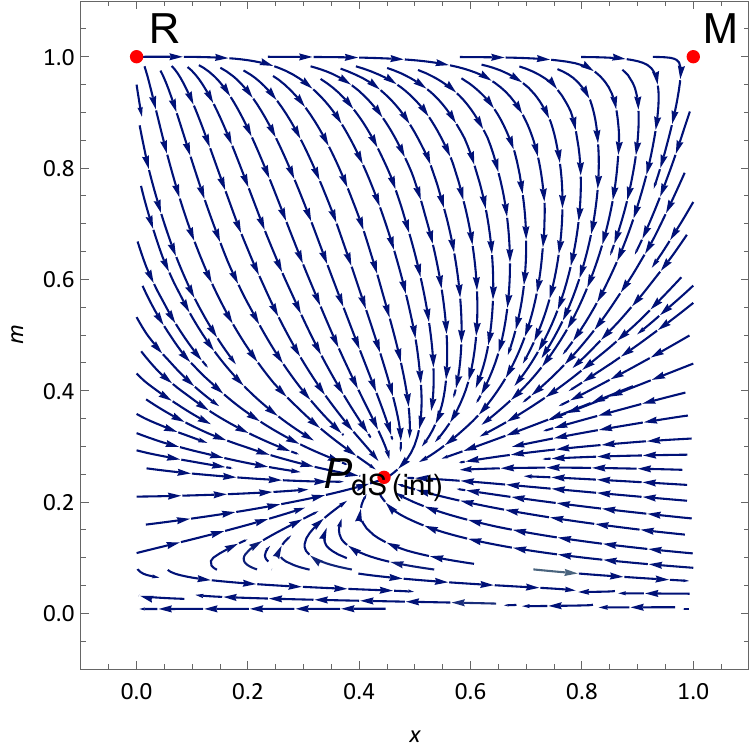} 
\caption{Phase space portrait for $\xi=3/2$ and $b=1/10$. The attractor point moves along the $x$ axis, signaling that matter component does not vanish in this hypothetical stage of cosmological evolution.}
\label{phasespacext}
\end{figure}

If we take larger positive values of the coupling parameter and $b>0$, the phase space plot changes more dramatically, since we have $x_{int}\neq 0$ at the de Sitter point, implying that there is no complete dominance of dark energy. However, qualitatively the de Sitter point still appears in the phase space as an attractor, driving the universe into a state of accelerated expansion in which dark matter does not completely dilute, presumably favoring the persistence of gravitational structures. This point is shown to illustrate the influence of the coupling parameter, but values greater than 1 are not statistically favored according to observations, as the interaction must remain very small \cite{Wang2024}.

\subsection{Stability analysis and cosmological parameters}

From the Jacobian matrices evaluated at the critical points $R$ and $M$, we obtain results similar to those in \ref{jacobians}. The changes in stability with respect to the previous Jacobian occur at the point $P_{int}$, where we obtain the functions $\mathbf{h}=\partial_{\mathbf{x}}\mathbf{f}'$, with $\mathbf{h}=\{h_{1}, ..., h_{4}\}$ forming a vector of functions derived from the partial derivative of the vector $\mathbf{f}'=\{f'_{1}, f_{2}\}$ with respect to the vector of dynamical variables $\mathbf{x}=\{x,m\}$, where $f'_{1}=f_{1}+\xi u(m)$. The Jacobians take the form
\begin{align}
J|_{R}=\begin{pmatrix}
1 & -\xi/3\\
0 & 4
\end{pmatrix}, \quad 
J|_{M}=\begin{pmatrix}
-1 & 4/3-\xi/3\\
0 & 3
\end{pmatrix}, \quad 
J|_{P_{int}}=\begin{pmatrix}
h_{1}(b,\xi) & h_{2}(b,\xi)\\
h_{3}(b,\xi) & h_{4}(b,\xi)
\end{pmatrix}.
\end{align}
where $h_{1}(b,\xi)$, $h_{2}(b,\xi)$, $h_{3}(b,\xi)$ and  $h_{4}(b,\xi)$ are given by
\begin{equation}
\begin{aligned}
h_1(b,\xi)=& -\dfrac{3\big[-b^{2} U\,S^{3}+14 b\,U\,S^{2}T-48\,U(48+6b+\xi)\,T^{2}+6\big(V(36-21\xi)+4U\big)\,T^{3}
\big]}{U\big[-54\,V\,T^{3}+b\,S^{2}\big(b(-12-\xi+2\sqrt{6}\,D)+6(12+\xi+2\sqrt{6}\,D)\big)\big]}
\end{aligned}
\end{equation}

\begin{equation}
\begin{aligned}
h_2(b,\xi)=&\frac{2\,\xi\,W}{27\,T^{4}}\Bigg\{6bV\,W\,S\,T-bV\,W\,S\big(bW-3(12+\xi+2\sqrt{6}\,D)\big)+3V\,S\,T\,R
+3V\,T^{2}R\\[4pt]
&+54\,T^{2}\Big[-b^{2}U\,S^{3}+14bU\,S^{2}T-48U\,W\,T^{2}+6\big(-9V^{-1}(4-\xi)+4U\big)T^{3}\Big]\\[4pt]
&-\;54\,T^{2}\Big[2b^{3}W\,S^{3}-2b^{2}U\,S^{2}T(780+192b+23\xi+18\sqrt{6}\,D)\\[2pt]
&+12bU\,S\,T^{2}(-228+192b+11\xi+42\sqrt{6}\,D)+36T^{3}\Big(16U W-3\big(-9V^{-1}(4-\xi)+4U\big)T\Big)\Big]\Bigg\}\\[6pt]
&\Bigg/\Big\{(12+\xi)^{2}\Big[ -54V\,T^{3}+bS^{2}\big(b(-12-\xi+2\sqrt{6}\,D)+6(12+\xi+2\sqrt{6}\,D)\big)\Big]^{2}\Bigg\},
\end{aligned}
\end{equation}

\begin{equation}
h_3(b,\xi)=\frac{27\,S\,T^{4}}{2(48+6b+\xi)^{2}\left[54\,T^{3}+b\,V\,S^{2}\big(b(12+\xi-2\sqrt6\,D)-6(12+\xi+2\sqrt6\,D)\big)\right]}
\end{equation}

\begin{equation}
\begin{aligned}
h_4(b,\xi)=& -\frac{12\,e^{-b/3}\,T_0}{U\,W\big[-54\,V\,T^{3}+b\,S^{2}\big(b(-12-\xi+2\sqrt{6}\,D)+6(12+\xi+2\sqrt{6}\,D)\big)\big]^{2}},
\end{aligned}
\end{equation}
with 
\begin{equation}
\begin{aligned}
T_0=&(T) \Big\{324\,T^{5}\big(E_{2}\,U\,S-72\,E_{3}\,R_{1}\big)[4pt]-3b^{2}S^{3}T\big(4E_{1}U\,K_{1}+6E_{2}T\,K_{2}\big)\\[4pt]
&+2b^{3}S^{4}\big(E_{1}U\,S^{2}-18E_{2}W\,T\big)
+9bS^{2}T^{2}\big(4E_{1}U\,S^{2}-6E_{2}T\,K_{3}\big)\Big\}
\end{aligned}
\end{equation}
and
\begin{equation}
\begin{aligned}
D &= \sqrt{6 - 36b + \xi},\quad S = 84 + \xi - 2\sqrt{6}\,D,\quad T= 12 + 12b + \xi + 2\sqrt{6}\,D,\\[4pt]
U &= 12 + \xi,\quad W = 48 + 6b + \xi,\quad V = \exp\!\left( \frac{b\,S}{6\,T} \right),\\[6pt]
R &=b(48+\xi-2\sqrt{6}\,D)-3(12+\xi+2\sqrt{6}\,D),\quad E_{1} = \exp\!\left( \frac{b}{3} \right),\quad E_{2} = \exp\!\left( \frac{b\,(108 + 24b + 3\xi + 2\sqrt{6}\,D)}{6\,T} \right),\\[6pt]
E_{3} &= \exp\!\left( \frac{2b\,W}{3\,T} \right),\quad K_{1} = 102 - 3b + \xi - 3\sqrt{6}\,D,\quad
K_{2} = -3312 - 108\xi - \xi^{2} - 6b(48 + \xi)
       + 24\sqrt{6}\,D,\\[4pt]
K_{3} &= 5184 + 5\xi^{2} + 6b(108 + 7\xi)
       + 4\xi\left(69 + \sqrt{6}\,D\right),\quad R_{1} = -18 + 3b + \sqrt{6}\,D.
\end{aligned}
\end{equation}

The eigenvalues of the Jacobians $J|_{R}$ and $J|_{M}$ remain the same, leading to the same stability for the radiation- and matter-dominated points, respectively. However, at the point $P_{int}$ the eigenvalues follow from the quadratic equation
\begin{align}
(h_{1}-\lambda_{int})(h_{4}-\lambda_{int})-h_{2}h_{3}=0,\\
\Rightarrow \lambda^2_{int}-\lambda_{int}(h_{1}+h_{4})-h_{2}h_{3}+h_1h_4=0.
\end{align}
This yields the solutions
\begin{align}
\lambda^{\pm}_{int}=\frac{(h_{1}+h_{4})\pm \sqrt{(h_{1}-h_{4})^2+4h_{2}h_{3}}}{2}.
\end{align}

\begin{figure}[h!]
\centering
\includegraphics[scale=0.59]{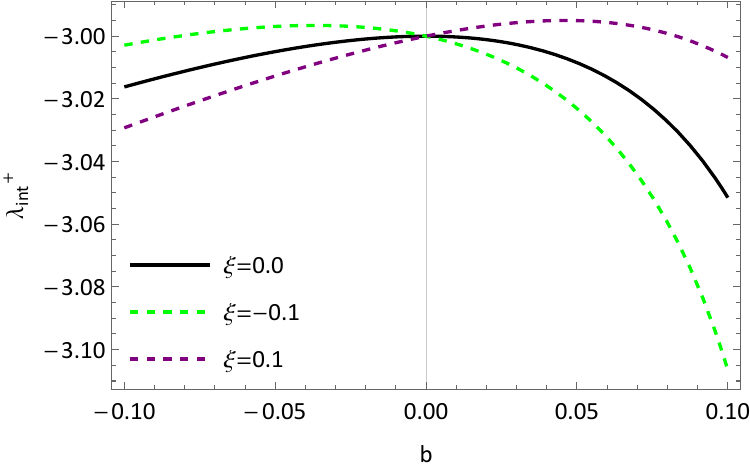} 
\hspace*{0.1 cm}
\includegraphics[scale=0.59]{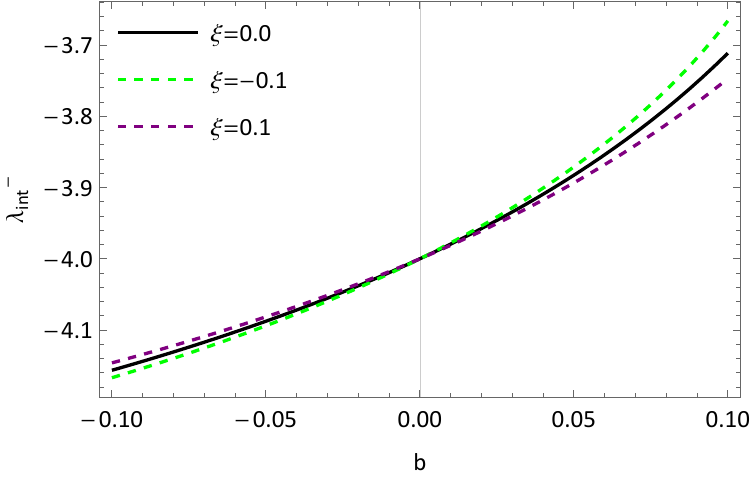}
\caption{Resultant eigenvalues as solutions to the quadratic equation, for different values of the coupling parameter $\xi$ and model parameter $b$.}
\label{eigenvalues1}
\end{figure}

Once again, due to the length of the expressions, we opted to plot the resulting eigenvalue functions. The figures (\ref{eigenvalues1}) show that, for varying parameters $(b,\xi)$, the eigenvalues $\{\lambda^{+}_{int}, \lambda^{-}_{int}\}$ associated with the point $P_{int}$ remain negative. Therefore, when the interaction is included, this equilibrium point is \textbf{stable}. Table \ref{tablesummarypoints} summarizes the critical points and their properties. It is clear that enlarging the parameter space of the model increases its complexity and may be statistically less favored compared to the $\Lambda$CDM model \cite{Wang2024}. Nevertheless, it is worthwhile to explore models without additional parameters, although this can be challenging since the interaction must remain sufficiently small, and a parameter such as $\xi$ allows one to tune the smallness of the interaction strength.

\begin{figure}[h!]
\centering
\includegraphics[scale=0.59]{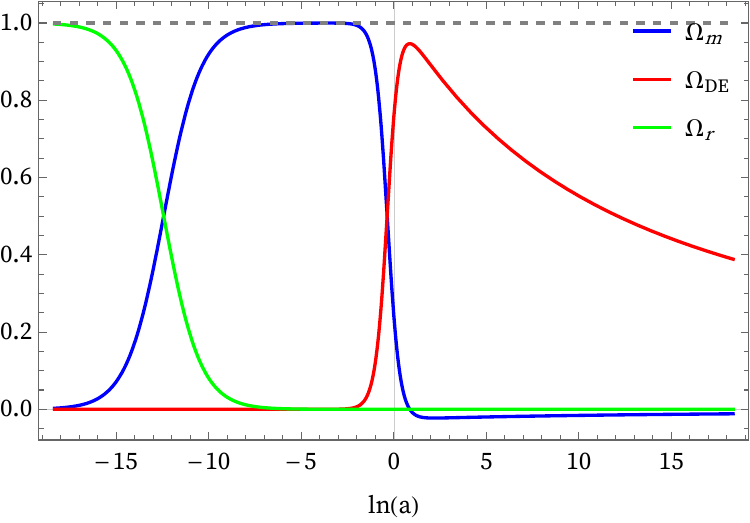} 
\hspace*{0.1 cm}
\includegraphics[scale=0.59]{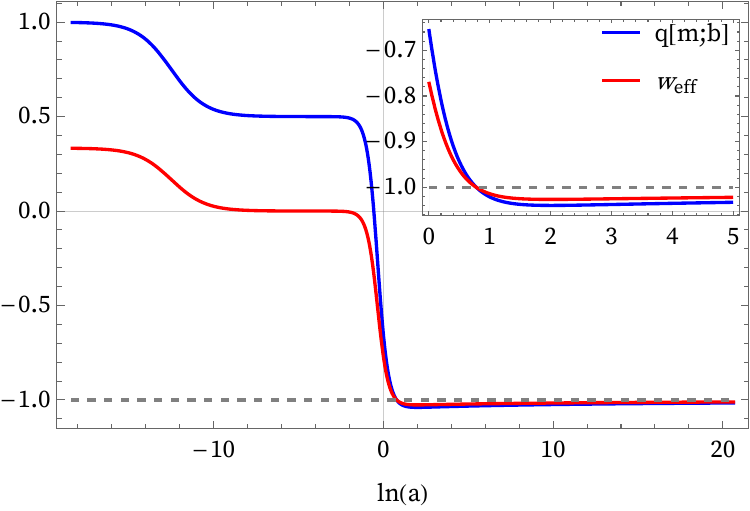}
\caption{Evolution of the state parameters with interaction, taking $\xi = -1/10$ and $b = -1/10$ [Left]. Evolution of the deceleration and effective equation-of-state parameters for the same values as before [Right].
}
\label{cosmopara}
\end{figure}

Finally, in the same way as before, we solve the autonomous system of equations with respect to the variable $\eta$ for the quantities $\{x,y,m\}$ in order to obtain the behavior of the corresponding density parameters. The equations are solved numerically, with the difference that the autonomous differential equation for $x$ now includes the contribution of the interaction:
\begin{align*}
\frac{dx}{d\eta}=-x\frac{[3t(m)-108m^3(3x/2+2y)]}{t(m)}+\xi u(m).
\end{align*}
We use the same initial conditions as before ($y(0)=10^{-6}$, $x(0)=1-\Omega_{DE}(b,0)-y(0)$, $m(0)=m_{0}$), and obtain the results shown in figures (\ref{cosmopara}). In this evolution of the density parameters in the presence of the interaction, we observe that the geometric component of dark energy does not completely dominate and decreases in value towards the future, while the matter component dissipates, taking negative values ($\Omega_{m}<0$) at the location of the dark-energy–dominated attractor point, a result which in principle is unphysical and will require more attention in further interacting models. It is interesting to note that this behavior is also generally found using the linear interaction kernel we chose to extend our model \cite{vander}. Moreover, at late times the effective equation-of-state parameter tends toward values $w_{eff}<-1$ and $q<-1$, suggesting the dominance of a phantom-like dark energy component. In this sense, it is interesting to note that the geometric contribution from non-metricity in the model can be decomposed into quintessence-like and phantom-like fields in the presence of interaction, since while the duality is controlled by the parameter $b$, the phantom divide is also crossed in the presence of the coupling parameter $\xi$. This phantom crossing posits a violation of the null energy condition \cite{Rubakov}, whose consequences in terms of how much of this violation can be accumulated in spacetime is bounded in semiclassical theories of gravity \cite{endure}.

\section{Summary and conclusions}\label{conclusions}

We have presented a general dynamical systems analysis for an exponential-type model in $f(Q)$ gravity, originally introduced by Oliveros and Acero~\cite{OliverosAcero2}. Choosing appropriate dynamical variables in modified gravity models is often challenging; therefore, we successfully employed the technique proposed by Böhmer \textit{et al.}~\cite{boehmerfq}, which allows the modified Friedmann equations to be reduced to a two-variable autonomous system. In our case, the variable $m$ was adapted to include the cosmological constant $\Lambda$, differing from the definition used in Böhmer’s work.

Applying the equilibrium condition to the system reveals three critical points: (i) a radiation-dominated point $R$, (ii) a matter-dominated point $M$, and (iii) a de Sitter attractor point $P_{dS}$. The deviation parameter $b$ affects the timing of the de Sitter epoch, producing an earlier onset for $b < 0$ and a delayed one for $b > 0$. Due to the complex nature of the resulting equations, the values of $m$ at the critical point $P_{dS}$ were determined using suitable approximations. Nevertheless, the analytical expressions obtained are consistent with the qualitative behavior observed in the phase-space portrait. Using linear stability theory, the critical points were characterized as follows: $R$ is an unstable point, $M$ is a saddle point, and $P_{dS}$ is an attractor, independently of the magnitude and sign of $b$. Consequently, the model behaves qualitatively like $\Lambda$CDM, as expected, since our proposed model represents a perturbative expansion around $\Lambda$CDM. Furthermore, the evolution of the density parameters exhibits the expected transitions from radiation to matter and finally to a dark-energy–dominated universe. For $b < 0$, we find $\Omega_{\mathrm{DE}} > 1$, implying that the dark energy density surpasses the critical density.

We also extended the model by introducing a linear interaction between dark energy (DE) and dark matter (DM). The interaction kernel was chosen to be relevant at late times, and we analyzed its effect on the evolution of the DE and DM densities. For DE, the coupling strength $\xi$ directly shifts the equation-of-state parameter $w_{\mathrm{DE}}$. We adopted $\xi < 0$, as preferred by recent observational constraints, which effectively drives a phantom crossing of $w_{\mathrm{DE}}$ at both early and late times. For DM, the same parameter primarily affects the density at the present epoch. As in the non-interacting case, the system exhibits three critical points that now depend on both parameters $b$ and $\xi$: (i) the radiation-dominated point $R$, (ii) the matter-dominated point $M$, and (iii) an interacting de Sitter attractor $P^{(\mathrm{int})}_{dS}$. For $\xi > 0$, the matter component does not vanish completely, while for $\xi < 0$, negative values of $\Omega_{m}$ appear. Using linear stability analysis, we again find $R$ to be an \textbf{unstable} point, $M$ a \textbf{saddle} point, and $P^{(\mathrm{int})}_{dS}$ a \textbf{stable} attractor, regardless of the specific values of $b$ and $\xi$. Finally, the evolution of the cosmological parameters indicates a violation of the null energy condition (NEC) in the far future, suggesting a dominant phantom-like contribution from the model.

In conclusion, the proposed exponential model proves viable in qualitatively describing all cosmological epochs of dominance. However, its functional form may become cumbersome when extended to include more general linear or nonlinear DM–DE interactions. Dynamical dark energy models appear to be well motivated by recent DESI results~\cite{desi2}, and interacting scenarios still have room within the dynamics of the so-called \textit{dark sector}. Nonetheless, both types of extensions to the standard $\Lambda$CDM model broaden the parameter space to be constrained, and without a fundamental guiding principle it remains difficult to reduce the number of free parameters. Modified gravity continues to be a plausible explanation for cosmic acceleration, yet forthcoming experiments and precision tests of General Relativity will be decisive in determining the future direction of such models. As future work, we plan to constrain the parameter space of our model using the latest DESI data releases.

\end{document}